# Evolution of Viral Pathogens Follows a Linear Order

**Zi Hian TAN**[a], **Kian Yan YONG**[a] and **Jian-Jun SHU**[a,✉]

---

[a] School of Mechanical & Aerospace Engineering, Nanyang Technological University, 50 Nanyang Avenue, Singapore 639798.
✉ Correspondence should be addressed to Jian-Jun SHU. E-mail address: mjjshu@ntu.edu.sg.

**ABSTRACT** Although lessons have been learned from previous severe acute respiratory syndrome (SARS) and Middle East respiratory syndrome (MERS) outbreaks, the rapid evolution of the viruses means that future outbreaks of a much larger scale are possible, as shown by the current coronavirus disease 2019 (COVID-19) outbreak. Therefore, it is necessary to better understand the evolution of coronaviruses as well as viruses in general. This study reports a comparative analysis of the amino acid usage within several key viral families and genera that are prone to triggering outbreaks, including coronavirus (SARS-CoV-2, SARS-CoV, MERS-CoV, HCoV-HKU1, HCoV-OC43, HCoV-NL63, HCoV-229E), influenza A (H1N1, H3N2), flavivirus (dengue virus serotypes 1-4, Zika) and ebolavirus (Zaire, Sudan, Bundibugyo ebolavirus). Our analysis reveals that the distribution of amino acid usage in the viral genome is constrained to follow a linear order, and the distribution remains closely related to the viral species within the family or genus. This constraint can be adapted to predict viral mutations and future variants of concern. By studying previous SARS and MERS outbreaks, we have adapted this naturally occurring pattern to determine that although pangolin plays a role in the outbreak of COVID-19, it may not be the sole agent as an intermediate animal. In addition to this study, our findings contribute to the understanding of viral mutations for subsequent development of vaccines and toward developing a model to determine the source of the outbreak.

**IMPORTANCE** This study reports a comparative analysis of amino acid usage within several key viral genera that are prone to triggering outbreaks. Interestingly, there is evidence that the amino acid usage within the viral genomes is not random but in a linear order.

In the past two decades, there have been three coronavirus outbreaks, with the current severe acute respiratory syndrome coronavirus 2 (SARS-CoV-2) causing a global pandemic[1–4]. The sequence of SARS-CoV-2 is distinct from severe acute respiratory syndrome coronavirus (SARS-CoV) and Middle East respiratory syndrome-related coronavirus (MERS-CoV) and shows a closer relationship with the coronavirus isolated in bat[5]. However, it is uncertain whether the virus was transmitted to human through an intermediate animal[6]. Currently, the long-lasting pandemic has caused a global death toll of 5.38 million and with variants constantly reduction vaccination effectiveness; there is no end in sight. In this study, we analyze the amino acid usage of all virus types and show that the viral gene-encoded amino acid usage exhibits a distinct pattern. Our findings reveal that several key viral families and genera that are prone to triggering outbreaks, including coronavirus (SARS-CoV-2, SARS-CoV, MERS-CoV, HCoV-HKU1, HCoV-OC43, HCoV-NL63, HCoV-229E), influenza A (H1N1, H3N2), flavivirus (dengue virus serotypes 1-4, Zika) and ebolavirus (Zaire, Sudan, Bundibugyo ebolavirus), have a constrained genomic variation pattern. Within the viral species, the distribution of amino acid usage is limited to a linear line and similar to each other. This constraint provides insight into the viral mutation mechanism and can be potentially useful for the prediction of future virus variants. Extending this finding by comparing with previous outbreaks of coronavirus, severe acute respiratory syndrome (SARS) and Middle East respiratory syndrome (MERS) and with the understanding of naturally occurring patterns, our findings can be applied in the early detection of the origin of the outbreaks.

**RESULTS**

*Amino acid usage pattern within viral genomes*

Here, it is demonstrated that the distribution function can be used to reveal the distinct pattern and characteristics of the viral genomes. Both exponential and quanta functions are defined by two parameters (a and b), which show a linear relationship when drawn for all virus coding sequences (CDS) (Figure 1). This suggests that the parameters do not have a random pattern. To find out whether these parameters have clinical significance, several representative strains of coronavirus, ebolavirus, flavivirus and influenza A group of viruses were selected to determine the distribution of their parameters. The CDS of each species was obtained through the filtering function of the National Center for Biotechnology Information (NCBI) virus database[7] and the National Institute of Allergy and Infectious Diseases Influenza Research Database[8]. Both exponential and quanta functions were similar; however, the quanta function was used for further analyses.

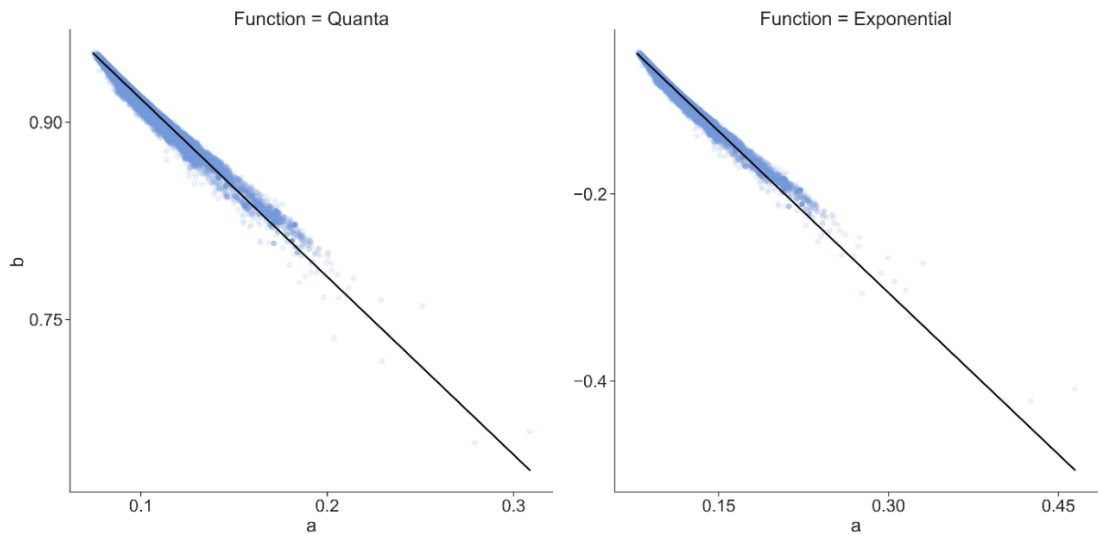

**Figure 1: All viral CDS a and b parameters from fitting exponential and quanta functions are represented as a scatterplot.** Each viral CDS with a fitted parameter is represented by a data point. The black diagonal line represents the linear regression of all data points. Overall, the viral CDS exponential and quanta function parameters show a linear relationship.

The parameters of viruses within the same family or genus are closely clustered together except for flavivirus, which exhibits a dispersed grouping (Figure 2). In the group of viruses, each species shows more distinct clustering of parameters, and most of the sequences lie on a straight line of ebolavirus, flavivirus and influenza A (Figure 3). Coronaviruses exhibit similar clustering characteristics (Figure 4). The sequence of SARS-CoV-2 is closely related to SARS-CoV, which is consistent with the findings of current pandemic viruses similar to the virus from the last SARS outbreak[5]. Of 2178 SARS-CoV-2 sequences, 2177 lie on the majority line along with other coronaviruses.

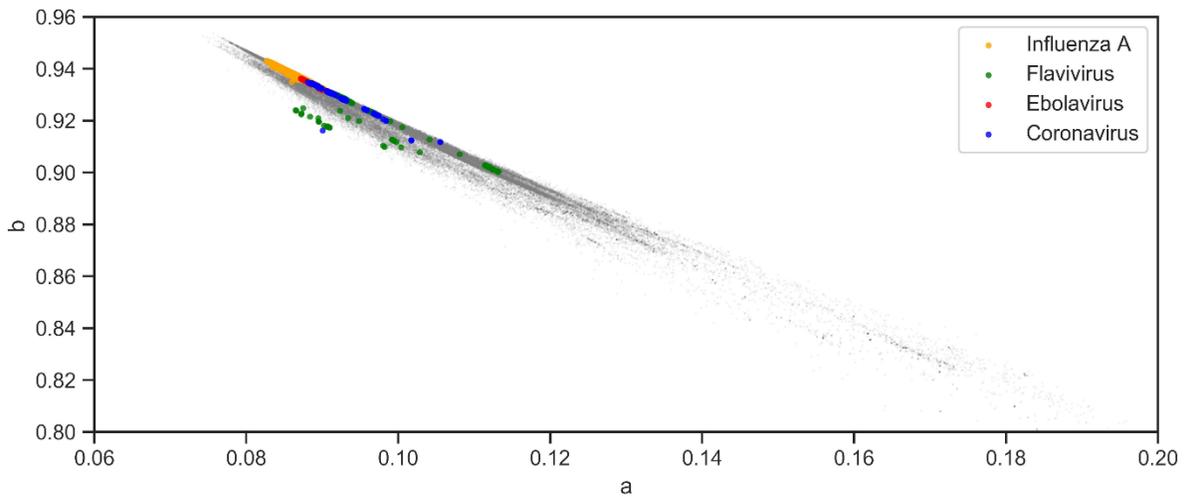

**Figure 2: Distribution of quanta parameters of four selected viral families in the context of all viruses (grey).** Representative viral species are influenza A (H1N1, H3N2), flavivirus (dengue virus serotypes 1-4, Zika), ebolavirus (Zaire, Sudan, Bundibugyo ebolavirus) and coronavirus (SARS-CoV-2, SARS-CoV, MERS-CoV, HCoV-HKU1, HCoV-OC43, HCoV-NL63, HCoV-229E).

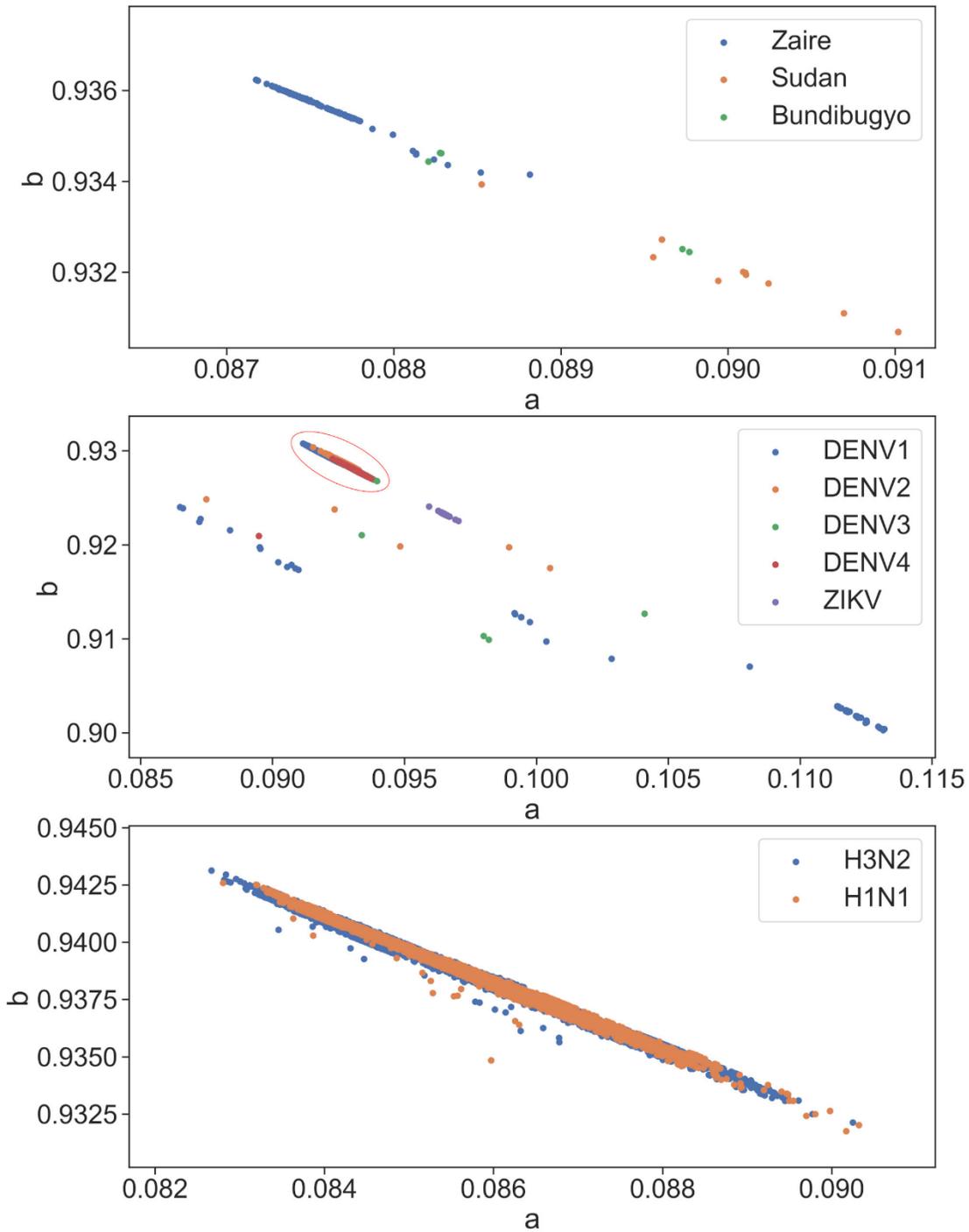

**Figure 3: Distribution of individual subgroups of viruses.** **(Top)** Ebolaviruses. Zaire strains have close distribution. Sudan and Bundibugyo strains with lesser sequences have a more dispersed distribution. **(Middle)** Flaviviruses. Most dengue viruses are clustered together (red circle) with serotypes 1-4 showing further clustering. Serotypes 1, 2 and 3 exhibit several outliers. These outliers follow a pattern of linear distribution and clustering. Zika

virus is clustered distinctly near the major dengue group. **(Bottom)** Influenza A. H1N1 and H3N2 do not exhibit a distinct cluster.

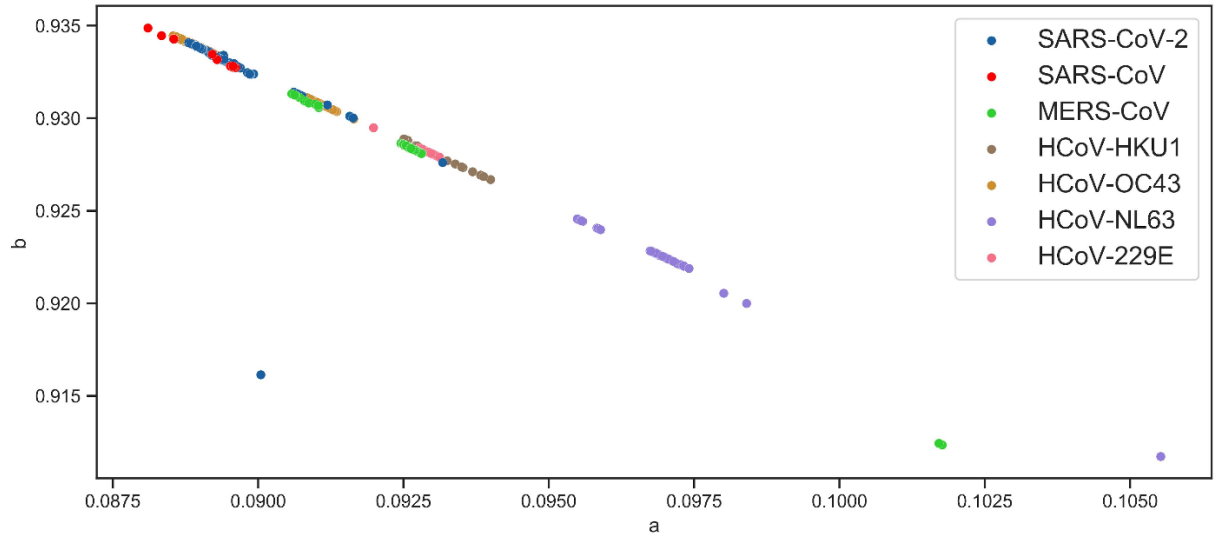

**Figure 4: Distribution of coronavirus subgroup.** Each species of coronavirus is grouped along a straight line with three (one SARS-CoV-2 and two MERS-CoV) outliers. Most of SARS-CoV-2 are clustered close to SARS-CoV.

*Prediction of intermediate host*

Most of the viral species show a nonrandom grouping of quanta distribution parameters. Next, the spike protein was analyzed to determine that it exhibits a similar pattern. The viral receptor is the most key component of the virus, as it determines the host infectivity and pathogenesis within the host. To explore the similarity of the spike protein parameters between the human coronavirus and the intermediate animal coronavirus, a quanta function was applied to the spike protein receptor sequence. The host standard was expanded to include all hosts from the NCBI virus database. The sequences of human and unknown hosts were filtered to form a zoonotic coronavirus reference for comparison with human coronavirus. The spike sequence of coronavirus was obtained by filtering from the NCBI database, except for SARS-CoV, as annotations are not available. The SARS-CoV spike sequence was obtained by gene alignment with the reference spike sequence (NC_004718.3) with the VIRULIGN software[9]. The spike sequence quanta parameters of the zoonotic 1656 coronaviruses were calculated with the same method as in the previous section. Three major coronaviruses, SARS-CoV-2, SARS-CoV, and MERS-CoV, were analyzed to illustrate that the spike protein distribution is more tightly grouped than the genome-wide CDS (Figure 5A). While it is expected for MERS-CoV to differ from SARS-CoV and SARS-CoV-2 due to the different target host receptor[10-12], the closely related coronavirus SARS-CoV and SARS-CoV-2 spike protein also show distinct grouping (Figure 5B). This is in contrast to the rest of the genome, where quanta parameters overlap (Figure 5C).

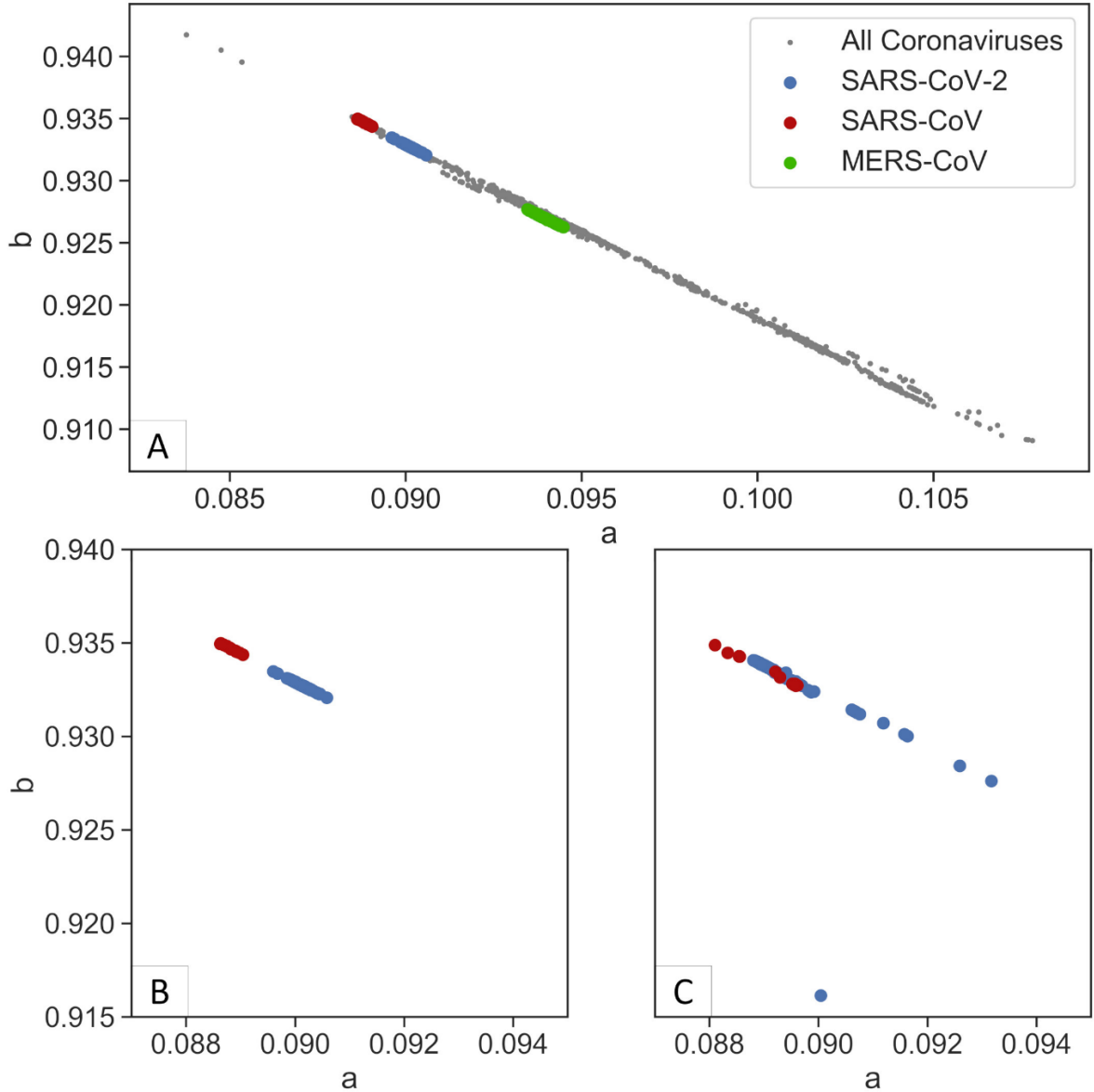

**Figure 5: Spike protein quanta distribution of SARS-CoV-2, SARS-CoV, and MERS-CoV. A.** Spike distribution for all coronaviruses from NCBI GenBank. The three major coronavirus spike proteins show distinct groupings. The SARS-CoV-2 spike protein cluster is closer to SARS-CoV than to MERS-CoV. **B.** The focused view of SARS-CoV-2 and SARS-CoV spike proteins shows a tighter and distinct clustering than the whole-genome CDS (**C**).

To determine whether this method can predict intermediate hosts, two coronaviruses that are responsible for previous outbreaks, SARS and MERS, were analyzed with zoonic hosts. The SARS-CoV spike protein shows clear clustering near the zoonotic coronaviruses of bat, civet, mouse, and grivet (Figure 6). Mouse (*Mus musculus*) and grivet/African green monkey (*Chlorocebus aethiops*) samples were experimentally infected[13]; to our knowledge, neither of these animals play a role in the SARS outbreak. The *Rhinolophus* genus of bats, which is identified as the closest match to SARS-CoV, is consistent with previous findings; that is, this genus is a natural reservoir host[13,14]. The intermediate animal in the SARS outbreak is identified as masked palm civet (*Paguma larvata*), but the zoonotic coronavirus sequences native to this species are unavailable. However, this model identifies other civet species (*Viverridae*, *Paradoxurus hermaphroditus*) as intermediate hosts by proximity. To quantify the nearest zoonotic coronavirus, the geometric distance (distance of the two points on the graph) between each of the 44 SARS-CoV proteins and each zoonotic coronavirus was determined (Figure 7). Bat species together with civets are ranked as the species closest to all SARS-CoV spike sequences. Interestingly, the spike protein of SARS-CoV shows the distinction between late and early midphase of the SARS epidemic, as noted in a previous study[15]. The early- and middle-phase sequences are grouped near to civets (*Viverridae* and *Paradoxurus hermaphroditus*) and bats; however, a clear migration upwards along the linear line toward mouse (*Mus musculus*) and grivet/African green monkey (*Chlorocebus aethiops*) is evident. The same calculation is performed for MERS-CoV (Figures 8 and 9) to show that camel is the closest host animal, and it is the previously identified intermediate host.

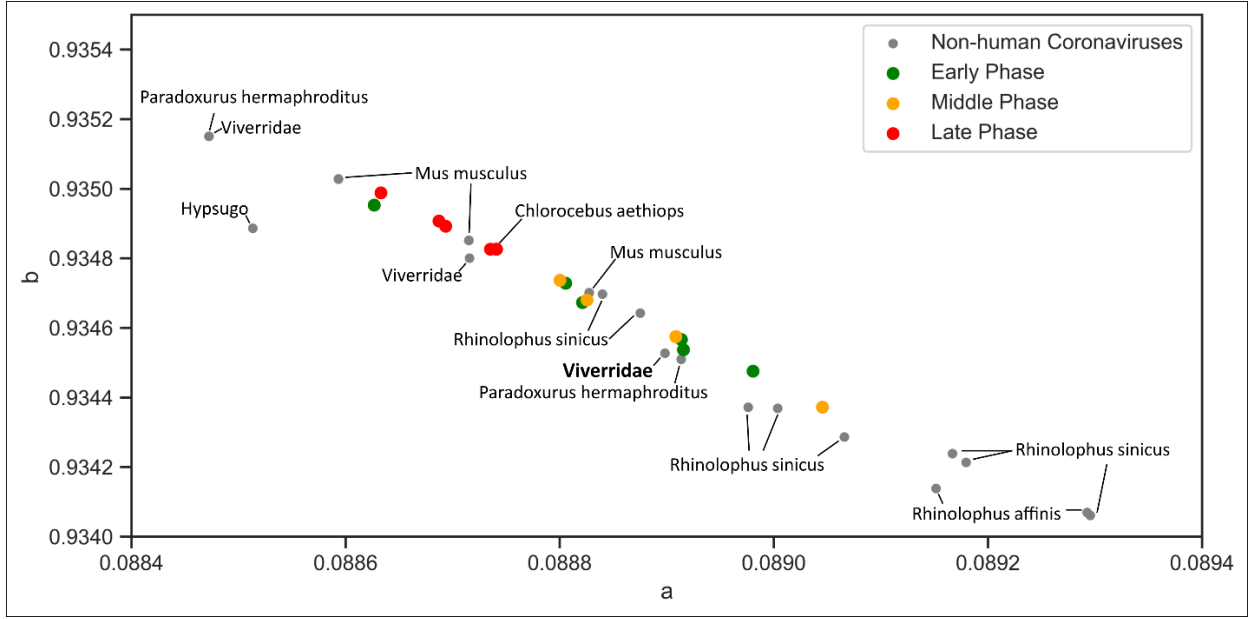

**Figure 6: Close-up view of SARS-CoV spike protein quanta parameter distribution.** SARS-CoV spike proteins are distributed near bat coronavirus spike proteins (*Rhinolophus*, *Hypsugo*), civet (*Viverridae*, *Paradoxurus hermaphroditus*), mouse (*Mus musculus*), and grivet (*Chlorocebus aethiops*). SARS-CoV quanta parameters are grouped into late (red), early (green), and middle phases (orange) of epidemic. The late-phase sequences are separated, except for the single early phase sequence.

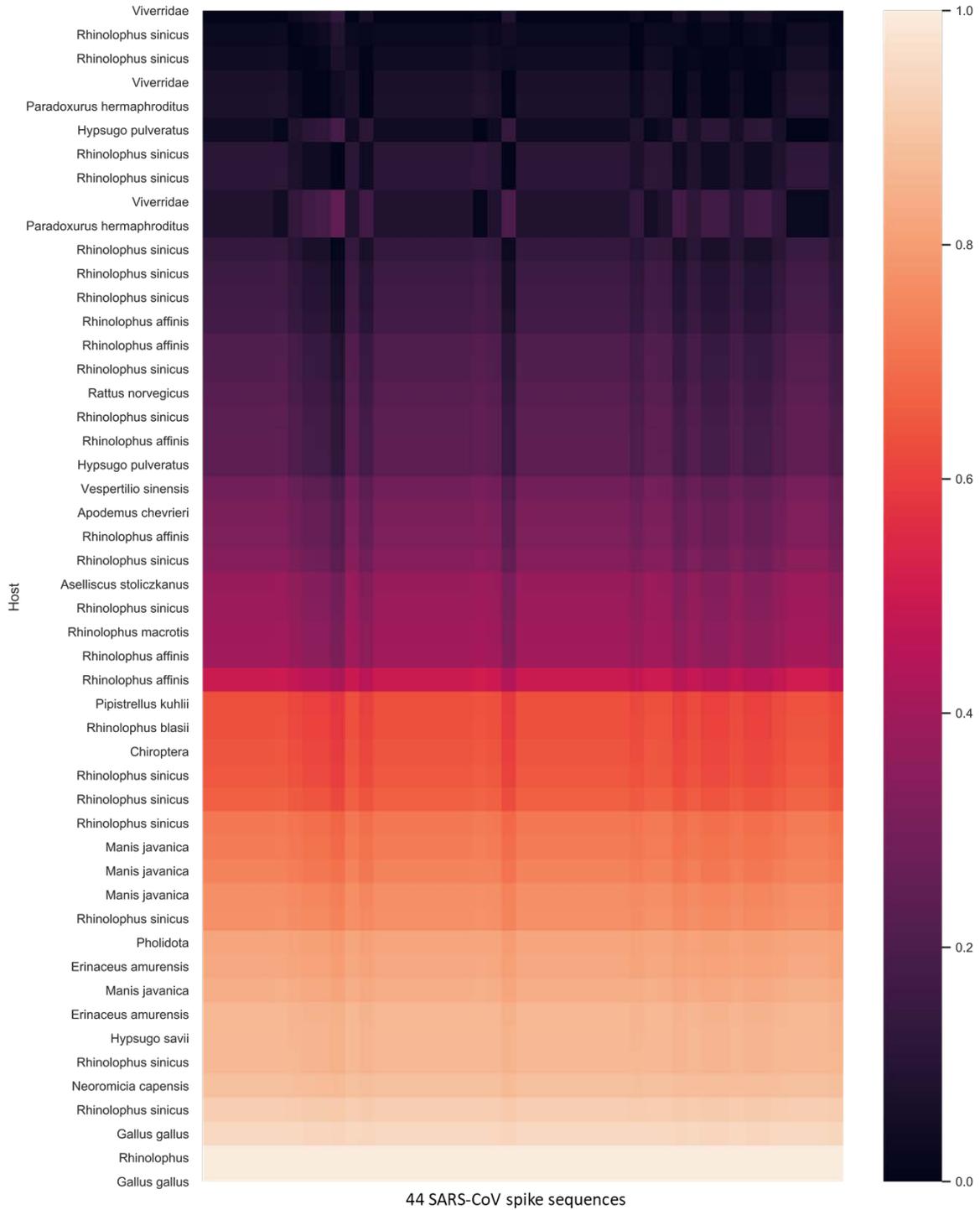

**Figure 7: Normalized geometric distance heat map of the nearest 50 of 1656 zoonic coronaviruses (vertical) and 44 SARS-CoV spike sequences (horizontal).** The heat map value represents the distance between human SARS-CoV and the zoonic coronavirus spike protein quanta parameter. The lower value (black) indicates that the zoonic coronavirus and human SARS-CoV spike quanta parameters are closest. The scale is normalized within the nearest

fifty samples. The majority of the near hosts are bat and small mammal groups, inclusive of the civet subfamily (*Viverridae*), Asian palm civet (*Paradoxurus hermaphroditus*), brown rat (*Rattus norvegicus*), and pangolin (*Manis javanica*, *Pholidota*). Mouse (*Mus musculus*) and grivet/African green monkey (*Chlorocebus aethiops*) are removed, as the viral samples are experimentally infected and do not represent natural hosts.

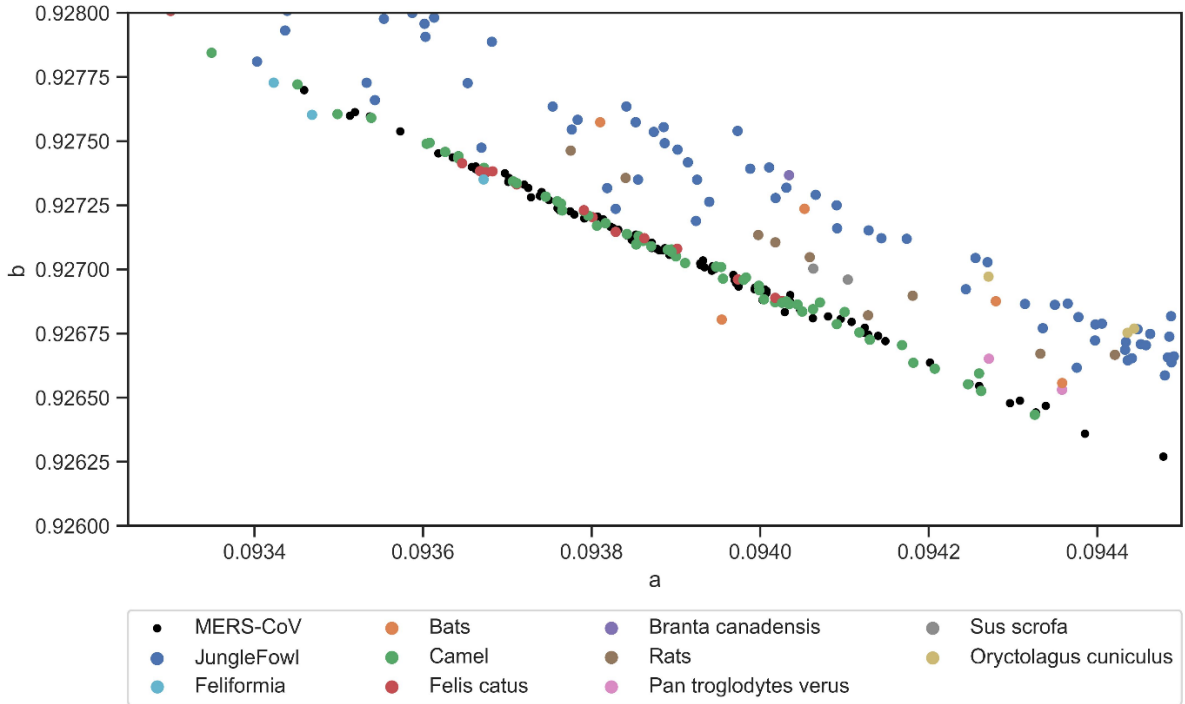

**Figure 8:** Close-up view of MERS-CoV spike protein quanta parameter distribution. MERS-CoV overlaps with camel coronavirus, indicating a close spike protein relation.

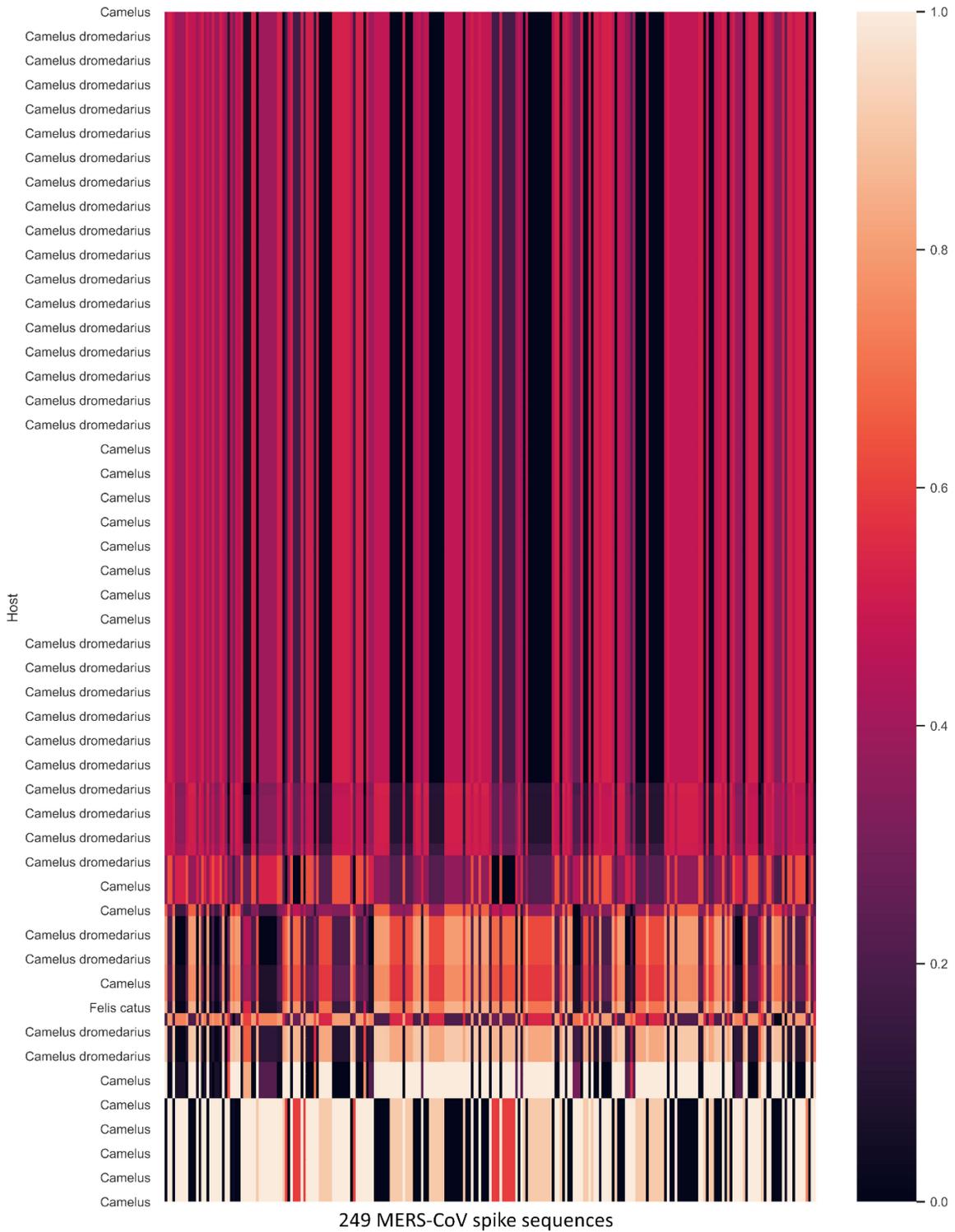

**Figure 9:** Normalized geometric distance heat map of the nearest 100 of 1656 nonhuman coronaviruses (vertical) and 249 MERS-CoV spike sequences (horizontal). *Camelus* (camel) coronavirus spike protein is closest to the MERS-CoV spike protein.

Using the same model validated against SARS-CoV and MERS-CoV, 1743 SARS-CoV-2 spike sequences available in the NCBI virus database were analyzed. Most of the sequences are distributed in the vicinity of bat coronaviruses except for eleven spike sequences (Figure 10), which are closer to Malayan pangolin. Two samples are from mainland China, one is from Taiwan, and the rest are from USA (Table 1). Geometric distance calculation reveals that the closest non-bat zoonotic host is the Malayan pangolin (Figure 11), which suggests that pangolin is an intermediate host or pangolin-CoV participates in the recombination of current SARS-CoV-2. This finding is consistent with two other studies[16,17], suggesting that Malayan pangolin plays a role in the current coronavirus disease 2019 (COVID-19) outbreak. Compared with the SARS-CoV distribution (Figure 6), the earliest SARS-CoV-2 spike sequence (NC_045512.2) was expected to be located near pangolin; however, it is located in the middle of the distribution. This indicates that the first reference SARS-CoV-2 sequence sampled from Wuhan is not the earliest strain if Malayan pangolin is the intermediate host.

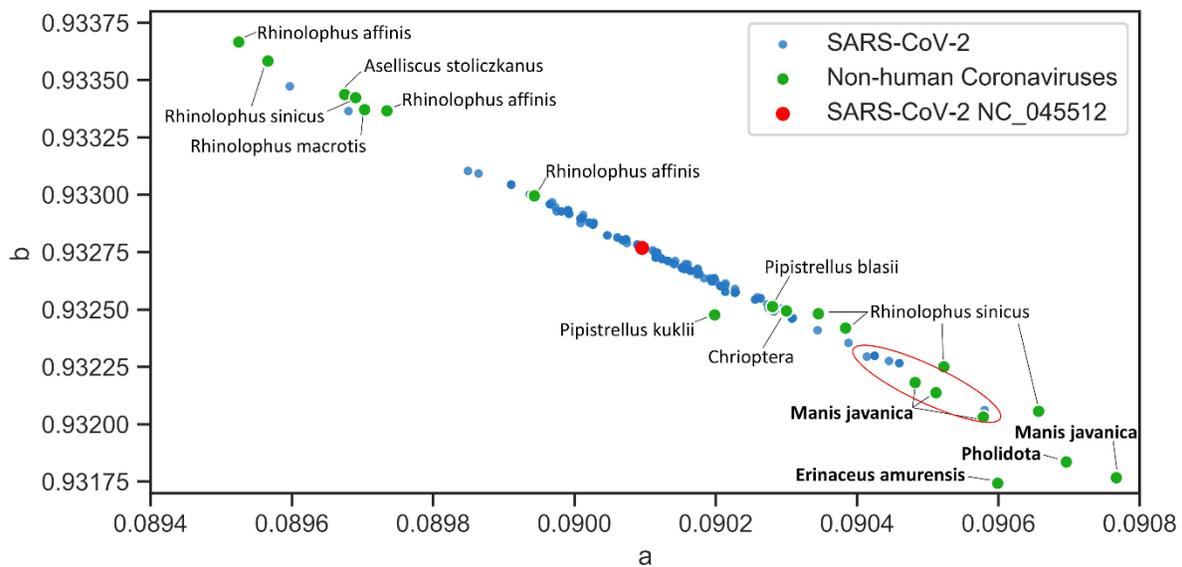

**Figure 10: Close-up view of spike sequence quanta parameter distribution centred on the SARS-CoV-2 reference genome (NC_045512.2).** The closest coronavirus host belongs to bats with the next nearest non-bat hosts

(bold), Malayan pangolin (*Manis javanica*), the general pangolin order (*Pholidota*), and Amur hedgehog (*Erinaceus amurensis*), as well as bat order (*Chiroptera*) and intermediate horseshoe bat (*Rhinolophus affinis*), vesper bat (*Pipistrellus kuhlii*), Chinese rufous horseshoe bat (*Rhinolophus sinicus*), big-eared horseshoe bat (*Rhinolophus macrotis*), and Stoliczka's trident bat (*Aselliscus stoliczkanus*).

**Table 1:** SARS-CoV-2 spike sequences near pangolin

| Accession | a | b | Location | Collection date |
|---|---|---|---|---|
| MT415371.1 | 0.09058 | 0.932063 | Mainland China | 2020-02-06 |
| MT334547.1 | 0.090459 | 0.932266 | USA: UT | 2020-03-19 |
| MT325615.1 | 0.090459 | 0.932266 | USA: IA | 2020-03-07 |
| MT374108.1 | 0.090445 | 0.932276 | Taiwan | 2020-03-14 |
| MT418890.1 | 0.090425 | 0.932298 | USA: VA | 2020-04 |
| MT418891.1 | 0.090425 | 0.932298 | USA: VA | 2020-04 |
| MT418892.1 | 0.090425 | 0.932298 | USA: VA | 2020-04 |
| MT418889.1 | 0.090425 | 0.932298 | USA: VA | 2020-04 |
| MT412303.1 | 0.090425 | 0.932298 | USA: CT | 2020-03-30 |
| MT385442.1 | 0.090425 | 0.932298 | USA: CA | 2020-04-03 |
| MT415370.1 | 0.090414 | 0.932295 | Mainland China | 2020-02-06 |

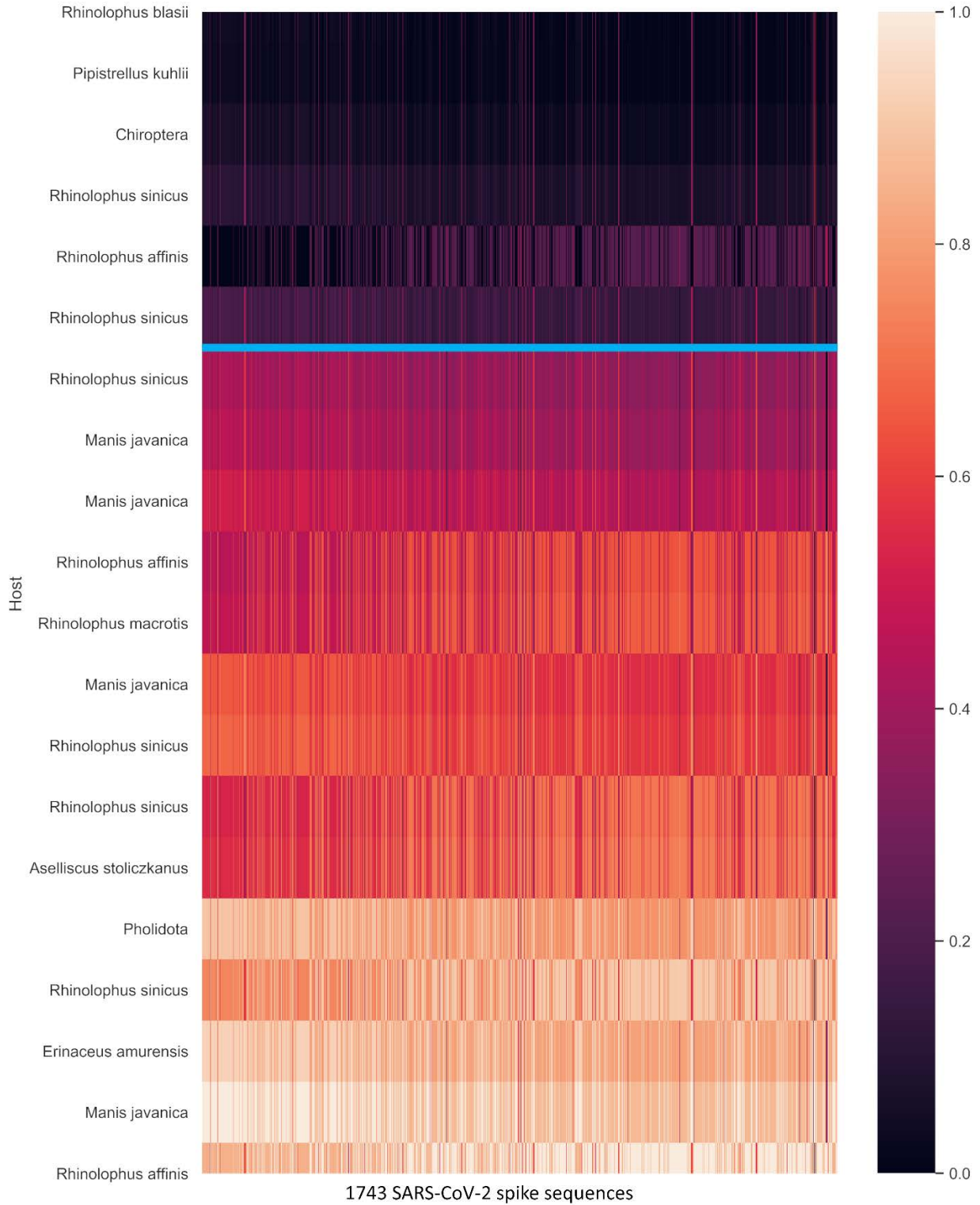

**Figure 11: Normalized geometric distance heat map of the nearest 20 of 1656 nonhuman coronaviruses (vertical) and 1743 SARS-CoV-2 spike sequences (horizontal).** The nearest host is bat (*Rhinolophus*, *Pipistrellus*,

*Aselliscus*), followed by pangolin (*Manis javanica*, *Pholidota*) and hedgehog (*Erinaceus amurensis*). The first six hosts are closer to SARS-CoV-2 (above blue line), and all six closest hosts are bats. Pangolin lies below the blue line.

**DISCUSSION**

Expectedly, each viral sequence differs in the distribution of amino acid, and thus the parameters describing the distribution vary. Yet, the distribution parameters are bounded by the linear relationship. The amino acid usage within the viral genomes is not random or does not follow stochastic behavior but correlates well to quanta and exponential distribution. The quanta distribution was selected for further analysis as it presents an interesting characteristic; that is, the highest rank holds the value of the parameter 'a', and each subsequent rank is lesser by a factor 'b'. The most frequently used amino acid has the usage percentage of 'a', and the second most frequently used amino acid has the usage percentage of 'a×b', and the next usage frequency is 'a×b×b'. Thus, the parameter 'b' that quantifies the decline or ratio of the next subsequent frequent amino acid is constant, which suggests that the amino acid usage of the virus follows a discrete and quantized nature. Hence, this distribution is termed quanta distribution. All viruses rely on the host's translation mechanism; hence, it is possible that this distribution exists in all living organisms' genomes. However, this requires further studies for confirmation.

One of the findings of this study is that the distribution parameters are not random across viral families and species. The quanta parameters of most viral sequences fall on a straight line. Distribution outliers can be caused by incomplete or noncurated sequences. Although these sequences are obtained from the NCBI virus database with the 'complete sequence' flag, a cursory examination reveals that the part of the sequence or CDS has incomplete proteins. It is an arduous task to manually curate all 115,285 sequences; however, most of the sequences exhibit a consistent linear characteristic with several outliers. Of the representative virus groups of interest (coronavirus, ebolavirus, flavivirus, and influenza A group), only flavivirus shows a significant amount of deviation from the linear line. The deviation is unlikely to arise from incomplete sequences. On closer analysis, the linear pattern is still consistent, and the flavivirus deviation forms parallel lines with the rest of the virus CDS (Figure 12). Flavivirus requires an insect vector for human transmission, and the dual-host requirement between a vertebrate and an arthropod places high selective pressure on flavivirus. It is known that flavivirus exhibits genomic plasticity and codon recoding under such pressures[18-20]. As this study focuses on human and zoonic coronavirus, the question of whether the dual-host requirement of flavivirus results in separate parallel distribution cannot be answered.

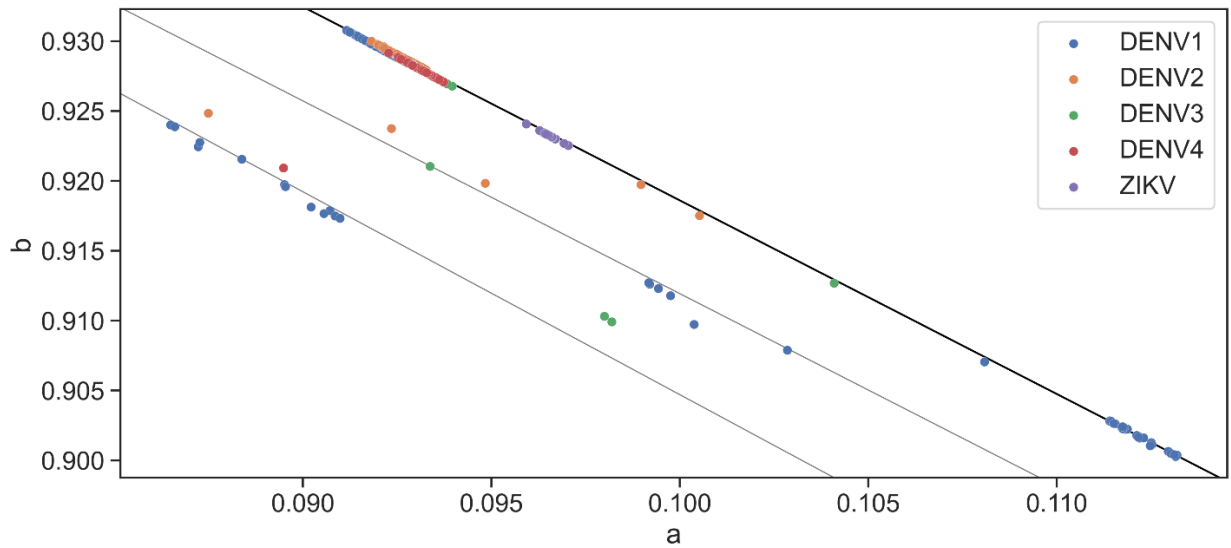

**Figure 12: Distribution of flavivirus subgroups.** Several sequences deviate from the main linear line (solid black). However, the deviation is not random but forms parallel lines (grey lines).

However, thus far, the results have shown that the viral coded genome has two constraints: the amino acid quanta distribution must fall on a linear line, and the resultant quanta parameters are closely grouped within the same viral species. This is especially true for coronavirus spike proteins. This implies that viral mutations do not occur randomly but are limited to the axis of a linear line, which is constrained in two directions and close to the original sequence. This characteristic of the viral genome can be used to predict future variants of SARS-CoV-2 and mutations in the spike protein used in vaccine development. For SARS-CoV-2, 2177 of the 2178 sequences show consistent distribution with other naturally evolved viruses. Therefore, the recent COVID-19 etiologic agent is likely to be the result of natural evolution.

Compared with the genome-wide CDS counterparts, the conservative distribution of spike proteins may be due to selective pressure, in which the fitness cost to receptor mutations is higher. This characteristic is adapted for predicting possible intermediate zoonotic hosts, and the findings are consistent with previously identified animals of SARS and MERS. The results of studies that pangolin may play a role in the current outbreak are consistent and support other studies[16,17]. Although more studies are needed to confirm the role of pangolin[16,17] or other possible animals in the current COVID-19 outbreak, this model can be able to rapidly identify the closest and most probable match for confirmation.

The SARS-CoV spike distribution shows the division between late and early midphase, and more case study is needed to correlate the phase of outbreak and the distribution of spike proteins. The SARS outbreak case study also shows the clear migration of late-phase spike protein quanta parameters toward mouse and grivet/African green monkey. It is unclear whether this migration is the result of adaptation to the urban environment or humans. Interestingly, the first sequence of the SARS-CoV-2 spike protein is located in the middle of the distribution contrary to SARS-CoV. It was expected that the spike sequence would be closer to the intermediate host of the initial root strain. There are two possibilities. First, if pangolin is the intermediate host of SARS-CoV-2, then the first sequence from Wuhan is not the root strain. The second possibility is that SARS-CoV-2 is likely to spread naturally from bats to humans without an intermediate host, which is likely to be adapted to pangolin at the same time. SARS-CoV and MERS-CoV are distributed near the intermediate hosts (Figures 6 and 8) and rank the closest quantitatively. For SARS-CoV-2, although pangolin is the next non-bat host, it is not the closest (Figure 11). Bat is obviously the closest to SARS-CoV-2, which leads us to consider the possibility that there may not be an intermediate host. Since bat is the natural reservoir for coronaviruses, and many have not crossed the human barrier, it is needed to consider the implications if more coronaviruses can spillover and reconsider our proximity from bats through the wildlife trade.

**CONCLUSIONS**

This study shows that virus mutation is not random but follows a set of constraints. Because of this constraint, it has the potential to be used for predicting future variants of concern and thereby speed up the development of specific and tailored vaccines. This method of predicting the closest animal host requires more case studies of previous outbreaks to establish a robust model framework. Although it cannot replace serological or other confirmation methods, infection and effective replication are not the only spike parameter similarity, but there are also other factors at play, such as the participation of receptor-binding sites or other nonstructural proteins. However, the findings have demonstrated the potential of using amino acid usage distribution as a rapid characterization tool for current and future novel virus outbreaks.

**MATERIALS AND METHODS**

To determine the overall pattern of amino acid diversity in all viruses, the CDS of all 115,285 viral sequences were obtained from GenBank and accessed through the NCBI virus database[7]. The amino acid frequency was obtained by the percentage of the amino acid count to the total number of coded amino acids in the sequence. The amino acid frequency of each complete CDS was sorted in descending order, independent of the order of amino acids, to eliminate the effect of codon bias.

As the order of amino acid is not constant, the amino acid frequency is defined as rank 1 to rank 20 in descending order. For the overall distribution pattern representative of all virus, the average of each rank was determined, and four (quanta $y = ab^{x-1}$, logarithmic $y = a\ln\left(1+\dfrac{b}{x}\right)$, power $y = ax^b$, and exponential $y = ae^{bx}$) functions that closely match the average overall distribution were selected (Figure 13). To identify which of the four functions best represents the usage of amino acids encoded within the viral genomes, each function was curve fitted to the amino acid usage distribution in each viral CDS, and the Pearson correlation coefficient for each fit was calculated. Among these four functions, the logarithmic and power functions fit with a mean correlation of 0.95, and the exponential and quanta functions fit with a mean correlation of 0.98. Among all viral sequences used in our analyses, the correlation coefficients of exponential and quanta functions have a low standard deviation of 0.012 (Table 2) and are highly concentrated near the median compared to logarithmic and power functions (Figure 14). For a universal function describing the amino acid distribution in all virus genomes, the function must be well correlated for every type of virus. Due to the high correlation, confidence ($P$ value), and narrow distribution of the curve fitting results for all virus sequences, the encoded amino acid usage within the viral genomes follows the exponential and quanta laws.

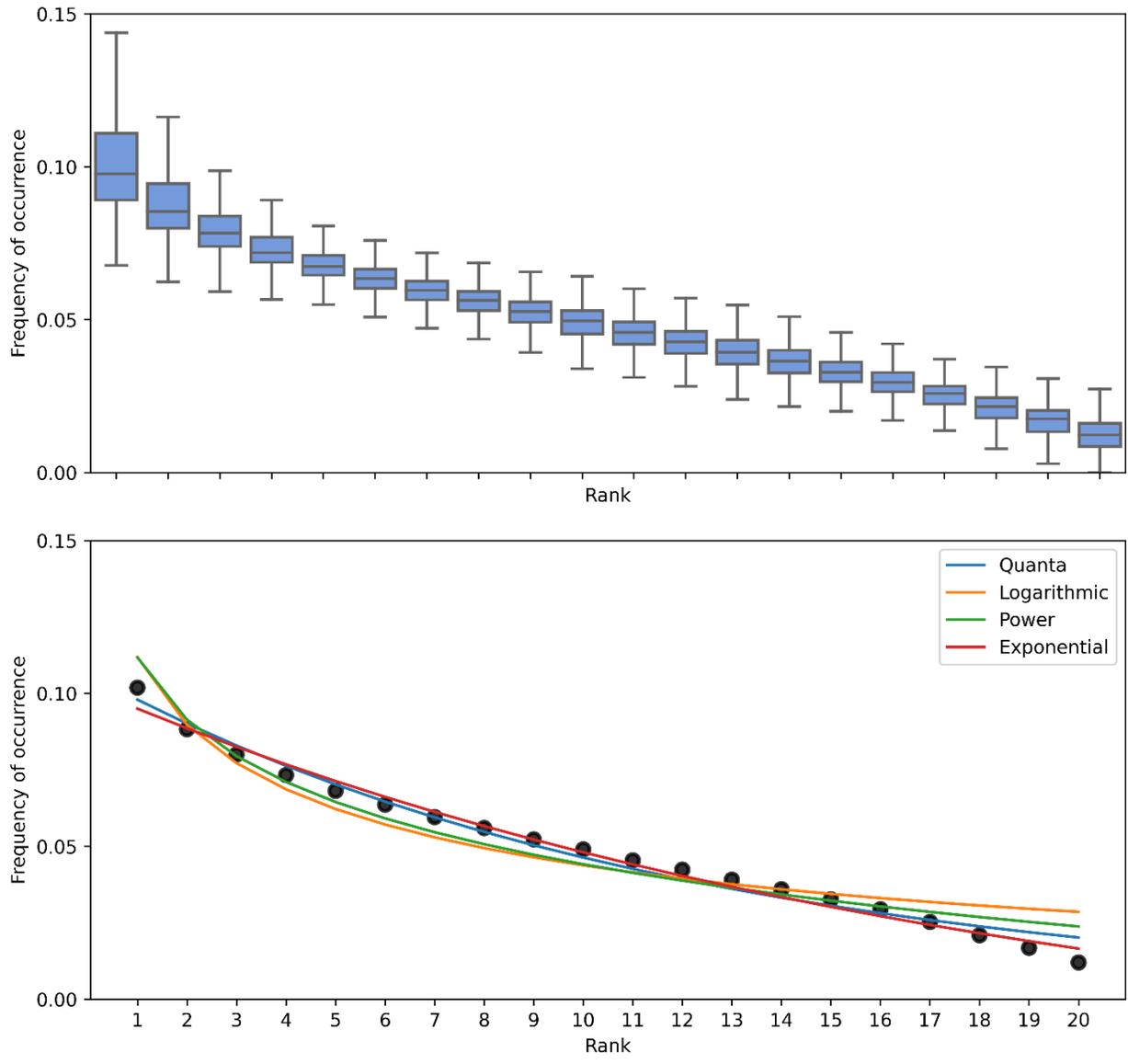

**Figure 13: Amino acid usage distribution of each viral CDS, excluding stop codons.** (Top) The amino acid usage distribution for each viral sequence was determined and represented as a percentage of the total amount of amino acid coded. The distribution was arranged from highest to lowest amino acid frequencies, represented by rank 1 to 20, respectively, as the amino acid preference of each virus differs. The distribution of every viral sequence at each rank is represented by a box plot. Rank 1 (most frequent amino acid) shows the largest amount of distribution, where ranks 4 to 20 are consistent. (Bottom) The black dots represent the mean frequency of each rank. The selected four (quanta, logarithmic, power, and exponential) curves fit the mean distribution.

**Table 2:** Summary of function fitting results

| Function | Pearson correlation | | | | Correlation $p$ value | | |
|---|---|---|---|---|---|---|---|
| | Standard deviation | Mean | Minimum | Maximum | Mean | Minimum | Maximum |
| **Exponential** | 0.012 | 0.98 | 0.86 | 1 | $2.5\times10^{-10}$ | $9.7\times10^{-27}$ | $10^{-6}$ |
| **Quanta** | 0.012 | 0.98 | 0.86 | 1 | $2.5\times10^{-10}$ | $9.7\times10^{-27}$ | $10^{-6}$ |
| **Logarithmic** | 0.037 | 0.95 | 0.74 | 1 | $1.6\times10^{-7}$ | $8.3\times10^{-24}$ | $1.7\times10^{-4}$ |
| **Power** | 0.034 | 0.95 | 0.76 | 1 | $8.6\times10^{-8}$ | $1.2\times10^{-23}$ | $1.1\times10^{-4}$ |

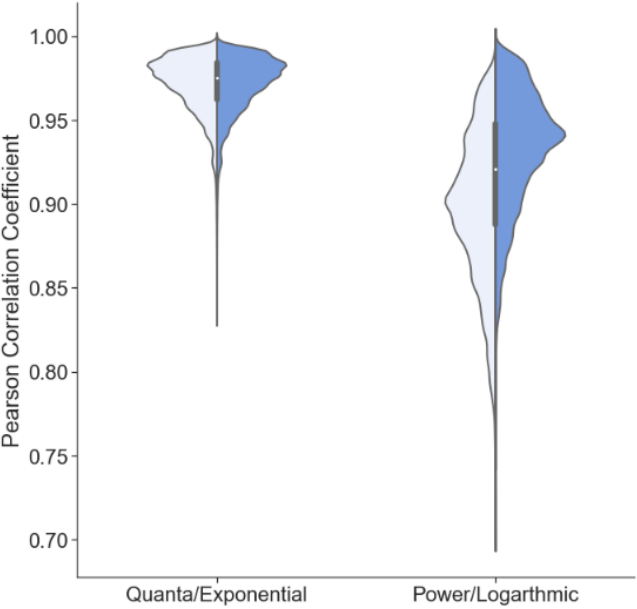

**Figure 14: Distribution of Pearson correlation coefficient values of each curve-fitting function to viral CDS.** Exponential and quanta functions exhibit similar distributions, with a narrow distribution (0.95-1.00) near the median value of 0.98. Logarithmic and power functions have a wider distribution spread out across 0.75-1.00. Exponential and quanta functions are a better and more consistent fit for all viral sequences.

**Declarations**

## Ethics approval and consent to participate

Not applicable.

## Consent for publication

Not applicable.

## Data availability

All data are available from the NCBI virus database and NIAID IRD.

## Competing interests

We declare no competing interests.

## Authors' contributions

All authors conceived, designed, and performed research, analyzed data, read, and approved the final manuscript.

## ACKNOWLEDGMENTS

This work was supported by Singapore Ministry of Education Academic Research Fund Tier 1 (04MNP002133C160).

2. Zhu N, Zhang DY, Wang WL, Li XW, Yang B, Song JD, Zhao X, Huang BY, Shi WF, Lu RJ, Niu PH, Zhan FX, Ma XJ, Wang DY, Xu WB, Wu GZ, Gao GF, Tan WJ. 2020. A novel coronavirus from patients with pneumonia in China, 2019. *New England Journal of Medicine* **382**(8):727–733.

3. Wu F, Zhao S, Yu B, Chen Y-M, Wang W, Song Z-G, Hu Y, Tao Z-W, Tian J-H, Pei Y-Y, Yuan M-L, Zhang Y-L, Dai F-H, Liu Y, Wang Q-M, Zheng J-J, Xu L, Holmes EC, Zhang Y-Z. 2020. A new coronavirus associated with human respiratory disease in China. *Nature* **579**(7798):265–269.

4. Zhou P, Yang X-L, Wang X-G, Hu B, Zhang L, Zhang W, Si H-R, Zhu Y, Li B, Huang C-L, Chen H-D, Chen J, Luo Y, Guo H, Jiang R-D, Liu M-Q, Chen Y, Shen X-R, Wang X, Zheng X-S, Zhao K, Chen Q-J, Deng F, Liu L-L, Yan B, Zhan F-X, Wang Y-Y, Xiao G-F, Shi Z-L. 2020. A pneumonia outbreak associated with a new coronavirus of probable bat origin. *Nature* **579**(7798):270–273.

5. Lu RJ, Zhao X, Li J, Niu PH, Yang B, Wu HL, Wang WL, Song H, Huang BY, Zhu N, Bi YH, Ma XJ, Zhan FX, Wang L, Hu T, Zhou H, Hu ZH, Zhou WM, Zhao L, Chen J, Meng Y, Wang J, Lin Y, Yuan JY, Xie ZH, Ma JM, Liu WJ, Wang DY, Xu WB, Holmes EC, Gao GF, Wu GZ, Chen WJ, Shi WF, Tan WJ. 2020. Genomic characterisation and epidemiology of 2019 novel coronavirus: Implications for virus origins and receptor binding. *Lancet* **395**(10224):565–574.

6. Liu ZX, Xiao X, Wei XL, Li J, Yang J, Tan HB, Zhu JY, Zhang QW, Wu JG, Liu L. 2020. Composition and divergence of coronavirus spike proteins and host ACE2 receptors predict potential intermediate hosts of SARS-CoV-2. *Journal of Medical Virology* **92**(6):595–601.
26